\documentclass{appolb}
\usepackage{epsfig}
\begin{document}
%
%
\def\koko{\mbox{K}_s^0\mbox{K}_s^0 }
\def\ra{\rightarrow }
\def\gam{\gamma }
\def\Ggg{\Gamma_{\gamma\gamma} }
\def\epem{\mbox{e}^+\mbox{e}^- }
\def\pip{\pi^+ }
\def\pim{\pi^- }
\def\ko{\mbox{K}^0 }
\def\kobar{\bar{\mbox{K}^0} }
\def\kol{\mbox{K}^0_L }
\def\kos{\mbox{K}^0_S }
\def\k{\mbox{K}}      
\def\pb{pb$^{-1}$}
\def\NP{{\it Nucl. Phys. }}
\def\PL{{\it Phys. Lett. }}
\def\ZfP{{\it Z. Phys. }}
\def\NIM{{\it Nucl. Inst. Meth. }}
\def\PRep{{\it Phys. Rep. }}
\def\PR{{\it Phys. Rev. }}
\def\PRL{{\it Phys. Rev. Lett. }}
\def\EURO{{\it Eur. Phys. J. }}
%
\pagestyle{plain}
\newcount\eLiNe\eLiNe=\inputlineno\advance\eLiNe by -1
\title{RESONANCE FORMATION\\
IN TWO-PHOTON COLLISIONS
}
\author{Saverio BRACCINI\thanks{Talk given at Meson2000, Cracow, Poland, May 2000.}
\address{University of Geneva,\\
24, Quai Ernest-Ansermet, CH-1211 Gen\`eve 4, Switzerland\\
{\it E-mail: Saverio.Braccini@cern.ch}}
}
\maketitle
\begin{abstract}
Two-photon collisions at the e$^+$e$^-$ colliders allow
to investigate the formation and the properties of resonant states 
in a very clean experimental environment.
A remarkable number of new results have
been recently obtained giving important contributions to meson
spectroscopy and glueball searches.
The most recent results from the LEP collider
at CERN and CESR at Cornell are reviewed here.
\end{abstract}

\section{Introduction}

 Two-photon collisions at electron positron storage rings are
a good laboratory to investigate the properties of meson
resonances and play a crucial role in glueball searches.
 
\begin{figure}[t]
\begin{minipage}[t]{60mm}
\mbox{\epsfig{file=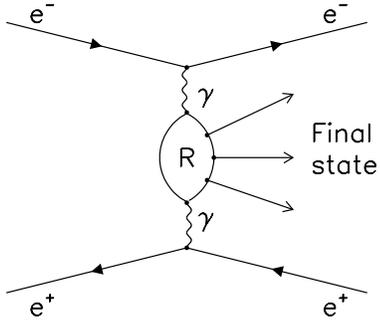,width=6.cm}}
\vspace*{-.4cm}
\caption{ Diagram of the formation of a resonant state in two-photon collisions
at e$^+$e$^-$ colliders.
}
\label{fig:ggr}
\end{minipage}
\hspace{\fill}
\begin{minipage}[t]{60mm}
\mbox{\epsfig{file=fig2.eps,width=6.cm}}
\vspace*{-.4cm}
\caption{The $\pi^+\pi^-\pi^0$ mass spectrum.
}
\label{fig:a2}
\end{minipage}
\vspace*{-.4cm}
\end{figure}

 A resonant state R can be formed by the collision of two photons via 
the reaction $\epem\ra\epem \mbox{R}$ (fig.~\ref{fig:ggr}). 
The outgoing electron and positron are usually scattered at very
small angles and are not detected (no-tag mode). In this case
the two photons are quasi real
and the resonant state R must be neutral and unflavoured with C=1 and
J$\neq$1. If one of the two photons is highly virtual, 
the outgoing electron or positron can be detected at low angle and the spin of
the resonant state is allowed to be one (single-tag mode). In both cases
the two outgoing particles carry nearly  the full 
beam energy and the mass of the resonant state is much smaller 
than the $\epem$ centre of mass energy.
This fact allows a clean separation between the two-photon and
the annihilation process by using a cut in the visible energy.
Since there are no particles produced other than R, 
the reconstruction of the final state can be 
performed in a very clean experimental environment.

 The cross section for this process
is given by the convolution of the QED calculable 
luminosity function $\cal{L}$, giving the flux of
the virtual photons, with the two-photon cross section
$\sigma(\gamma\gamma\ra\mbox{R})$ that is expressed by
the Breit-Wigner function 
\begin{equation}
\sigma
(\gamma\gamma\ra\mbox{R})
= 8 \pi (2\mbox{J}+1) 
\frac{\Gamma_{\gamma\gamma}(\mbox{R})\Gamma(\mbox{R})}
{(W_{\gamma \gamma}^2-m_R^2)^2+m_R^2\Gamma^2(\mbox{R})}
\label{eq:sgg}  
\end{equation}
where $W_{\gamma\gamma}$ is the invariant mass of the two-photon
system, $m_R$, J, $\Gamma_{\gamma\gamma}(\mbox{R})$
and $\Gamma($R)
are the
mass, spin, two-photon partial width 
and total width of R, respectively.
This leads to the
proportionality relation 
\begin{equation}
\sigma(\epem\ra\epem\mbox{R})=\mbox{$\cal{K}$}\cdot\Gamma_{\gamma \gamma}(\mbox{R})
\end{equation}
that allows to extract the two-photon width from the cross section.
The proportionality factor $\cal{K}$ is evaluated by a Monte Carlo integration.

In the single tag-mode the high virtuality of one of the two photons is
taken into account by multiplying the Breit-Wigner function by a VDM pole
transition form factor  
\begin{equation}
F^2(Q^2) = \left(\frac{1}{1+Q^2 / \Lambda^2} \right)^2 
\end{equation}
where 
$Q^2$ is the four vector squared of the virtual photon
and $\Lambda$ is a parameter to be measured experimentally.

 Since gluons do not couple directly to photons, 
the two photon width of a glueball is expected to be
very small. A state that can be formed in a gluon rich environment but
not in two photon fusion has the typical signature of a glueball.
According to lattice QCD predictions~\cite{LatticeQCD}, the ground state glueball has
J$^{PC}$= 0$^{++}$ and a mass between 1400 and 1800 MeV. 
The 2$^{++}$ tensor glueball is expected in the mass region around 2200 MeV
while the 0$^{-+}$ pseudoscalar glueball is predicted to be heavier.
Since several 0$^{++}$ states have been observed in the 1400-1800 MeV
mass region, the scalar ground state glueball can  mix with nearby quarkonia,
making the search for the scalar glueball and the interpretation
of the scalar meson nonet a complex
problem~\cite{Amsler}~\cite{Minkowski}~\cite{Gastaldi}.
The results from two-photon formation represent therefore a fundamental piece
of information for glueball searches.

 In order to distinguish ordinary quarkonia from glueballs,
a parameter called stickiness has been introduced~\cite{Chanowitz}.
The stickiness is an estimate of the ratio 
$|<R|gg>|^2 / |<R|\gamma\gamma>|^2$
evaluated from the 
the ratio 
$\Gamma(J/\psi\rightarrow\gamma R) / \Gamma(R\rightarrow\gamma\gamma)$
corrected by a phase space factor. The $s\bar{s}$ mesons have the
largest stickiness among quarkonia (14.7 for the f$_2'$(1525)) while much larger values 
are expected for glueballs.

 Because of the large mass of the charm quark, the study 
of the formation of charmonium states allows to test
non-relativistic perturbative QCD calculations and to measure $\alpha_s$ at the charm scale.
 
 Two e$^+$e$^-$ colliders have collected a large amount of data in
the last few years. The four LEP experiments ALEPH, DELPHI, L3 and OPAL
at CERN have collected approximately
150, 55, 175, 240 pb$^{-1}$ each at $\sqrt{s} \sim $ 91, 183, 189, 191-202  GeV 
respectively.
The CLEO experiment at
CESR (Cornell) has collected about 3000 pb$^{-1}$
at  $\sqrt{s} \sim $ 10.6 GeV.
Since the luminosity function $\cal{L}$ increases with the beam energy, 
the higher energy allows LEP to partially compensate the smaller luminosity by
a larger cross section.

 In this paper the most recent results on resonance formation and glueball
searches obtained at LEP and CESR are reviewed.

\section{The $\pi^+\pi^-\pi^0$ final state}

A study of the reaction $\gamma\gamma\rightarrow\pi^+\pi^-\pi^0$
is performed by L3~\cite{L3-a2}~\cite{Saverio-Frascati} using only untagged events.
The mass spectrum (fig.~\ref{fig:a2})
is dominated by the formation of
the a$_2$(1320) tensor meson. A clear enhancement is visible around 1750 MeV where the
study of the total transverse momentum distribution shows evidence for an exclusive process. 
The study of the angular distributions shows that the a$_2$ 
formation is dominated by a J$^{PC}$=2$^{++}$ helicity 2 wave. 
The radiative width is found to be
$\Gamma_{\gamma\gamma}(\mbox{a}_2)=0.98\pm 0.05 \pm 0.09$ keV. 
A spin-parity analysis
in the mass region above the a$_2$(1320) shows that also this region is 
dominated by a J$^{PC}$=2$^{++}$ helicity 2 wave, confirming 
the observation of the CERN-IHEP collaboration~\cite{Serpukov}
and in contradiction with CELLO~\cite{Cello} and Crystal Ball~\cite{CB} measurements.
This can be interpreted as the formation of a radial recurrence of the a$_2$ for 
which
$\Gamma_{\gamma\gamma}(\mbox{a}'_2(1765))\times$BR$(\mbox{a}'_2(1765)\rightarrow\pi^+\pi^-\pi^0)=0.29\pm 0.04 \pm 0.02$ keV 
in agreement with 
theoretical predictions~\cite{Munz}.
The J$^{PC}$=2$^{-+}$ wave contribution is found compatible with zero.

\section{Pseudoscalar mesons and their form factors}

\begin{figure}[t]
\begin{minipage}[t]{60mm}
\mbox{\epsfig{file=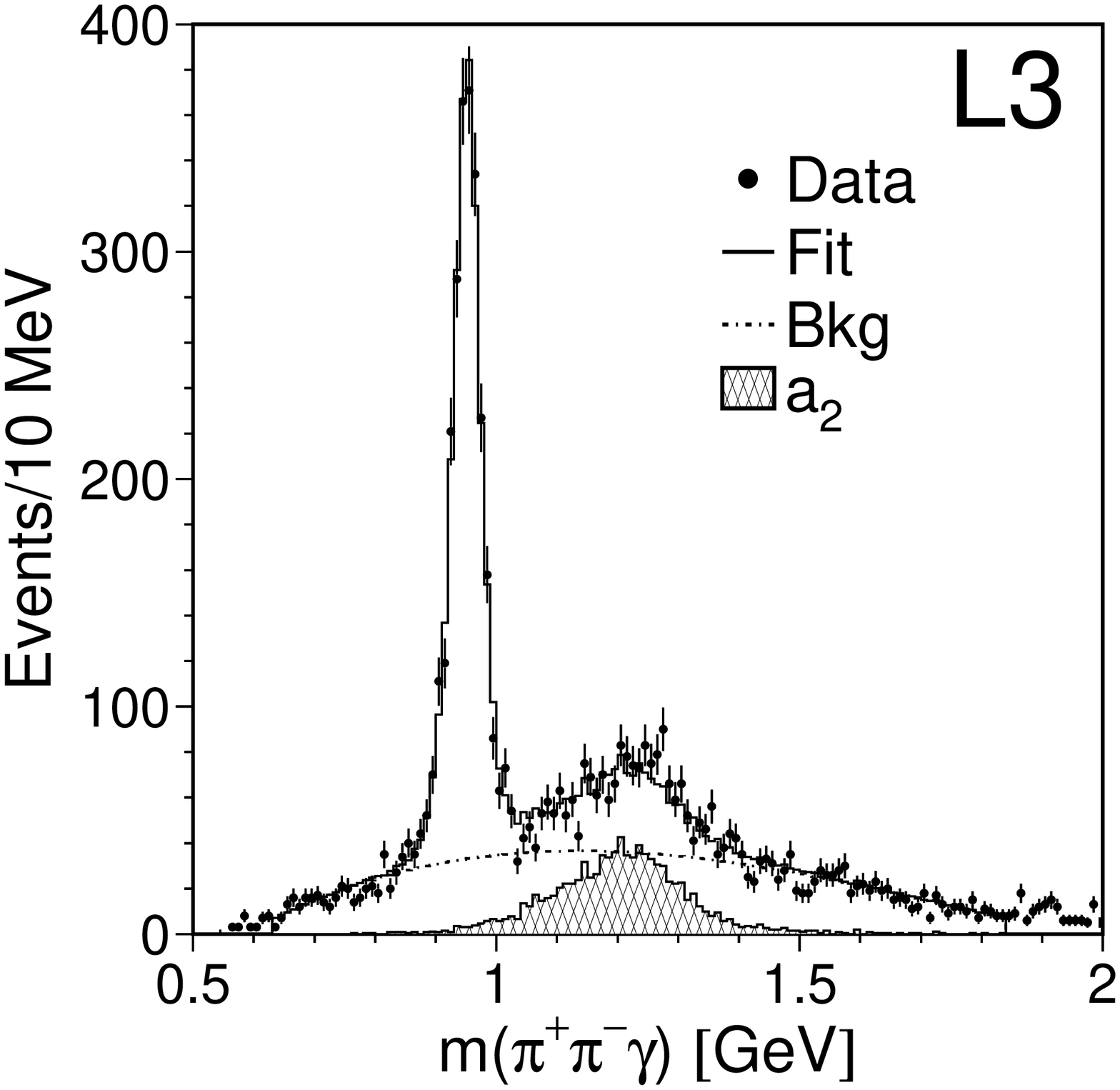,width=6.cm}}
\vspace*{-.4cm}
\caption{The $\pi^+\pi^-\gamma$ mass spectrum for $Q^2<0.01$ GeV$^2$.
}
\label{fig:etap}
\end{minipage}
\vspace*{-.4cm}
\hspace{\fill}
\begin{minipage}[t]{60mm}
\mbox{\epsfig{file=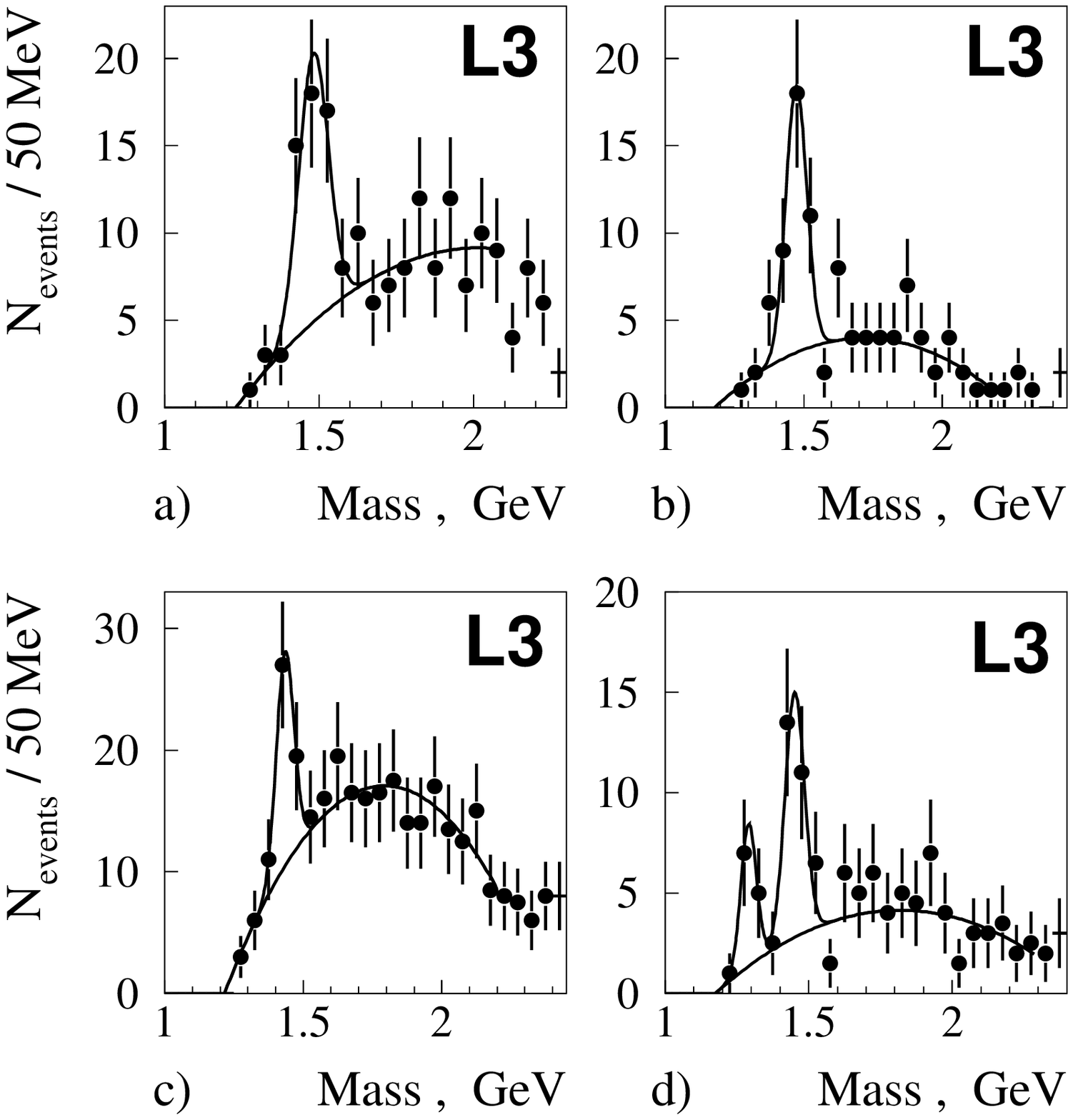,width=6.cm}}
\vspace*{-.4cm}
\caption{The K$_s^0$K$^\pm\pi^\mp$ mass spectrum for $Q^2<$0.02 GeV$^2$ (a),
0.02$<Q^2<$0.2 GeV$^2$ (b), 0.2$<Q^2<$1.0 GeV$^2$ (c) and 1.0$<Q^2<$7.0 GeV$^2$ (d). 
}
\label{fig:k0kp}
\end{minipage}
\end{figure}

 The reaction $\gamma\gamma\rightarrow\eta'\rightarrow\pi^+\pi^-\gamma$ is studied
by L3~\cite{L3-etap} in both the no-tag and single-tag mode. The $\pi^+\pi^-\gamma$
mass spectrum (fig.~\ref{fig:etap}) shows a prominent peak due to the formation of the
$\eta'(958)$ while the enhancement around 1250 MeV is due to the process
$\gamma\gamma\rightarrow$a$_2$(1320)$\rightarrow\pi^+\pi^-\pi^0$ when one photon 
from the $\pi^0$ goes undetected.
For the two-photon width,  
$\Gamma_{\gamma\gamma}(\eta')=4.17\pm 0.10 \pm 0.27$ keV is measured.
The electromagnetic form factor of the $\eta'$ is studied using tagged and 
untagged events. For the untagged events $Q^2$ can be measured 
as $(\Sigma p_t)^2$, as demonstrated by a Monte Carlo study.
A low gluonic component in the $\eta'(958)$ is found by
comparing the data with the theoretical predictions~\cite{Anisovitch-ff}. 
The value 
$0.900\pm0.046\pm0.022$ GeV is obtained for the parameter $\Lambda$.

 The transition form factors for the three pseudoscalar mesons $\pi^0$, $\eta$ and
$\eta'$ are studied by CLEO~\cite{CLEO-ff} using only the single-tag mode. The values
$\Lambda_{\pi^0}$=0.776$\pm$0.010$\pm$0.012$\pm$0.016 GeV,
$\Lambda_{\eta }$=0.774$\pm$0.011$\pm$0.016$\pm$0.022 GeV,
$\Lambda_{\eta'}$=0.859$\pm$0.009$\pm$0.018$\pm$0.020 GeV
are measured. Data are consistent with a similar wave function for the $\pi^0$
and $\eta$. The non-perturbative properties of the  $\eta'(958)$ are found to be different
from those of the $\pi^0$ and $\eta$. According to T. Feldmann~\cite{Feldmann},
another interpretation of these results leads to the conclusion that  
$\pi^0$, $\eta'$ and $\eta'$ mesons behave
similarly in hard exclusive reactions.

\begin{figure}[t]
\vspace*{-1.cm}
\includegraphics[width=0.45\textwidth]{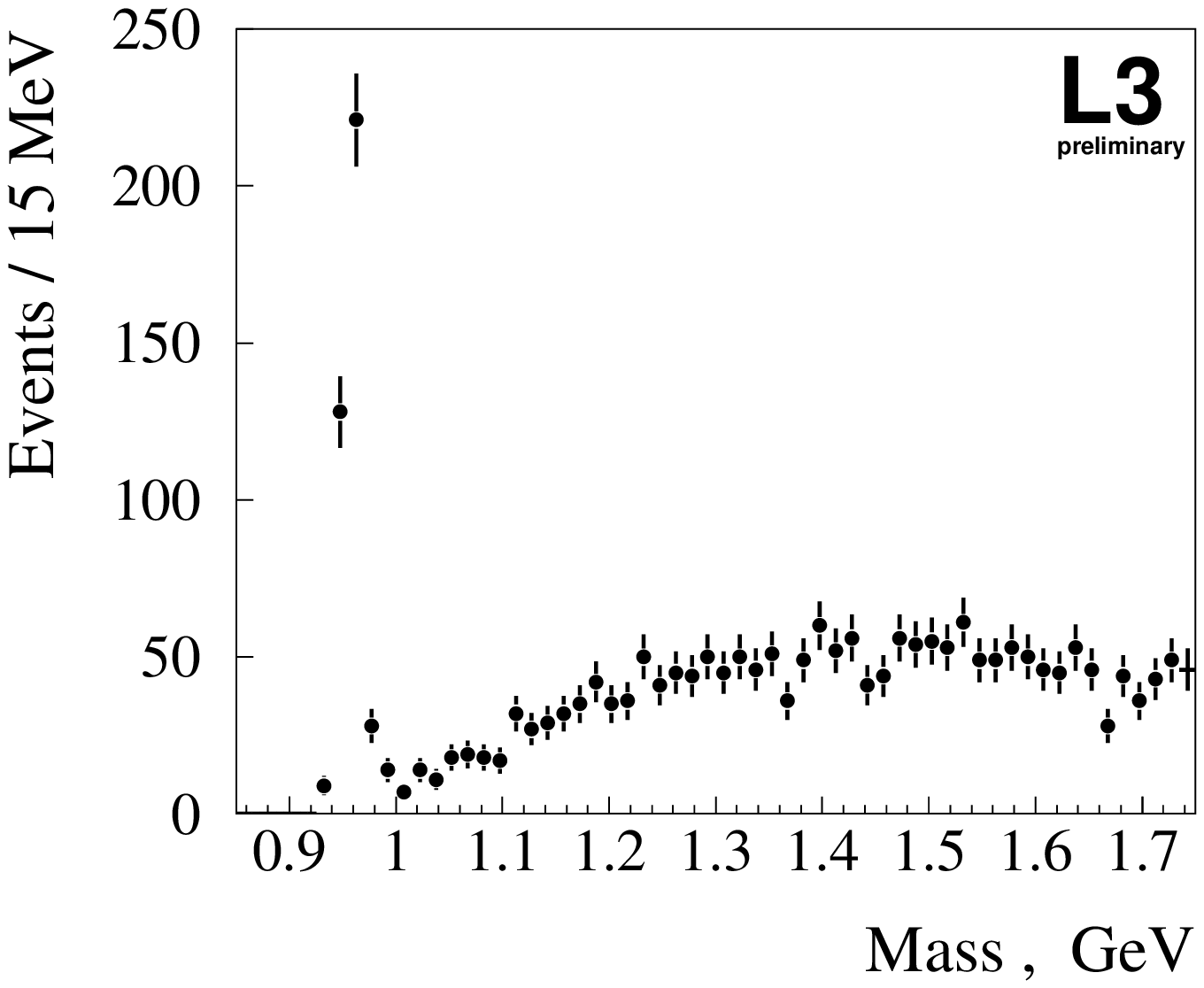}\hspace*{.3cm}
\includegraphics[width=0.45\textwidth]{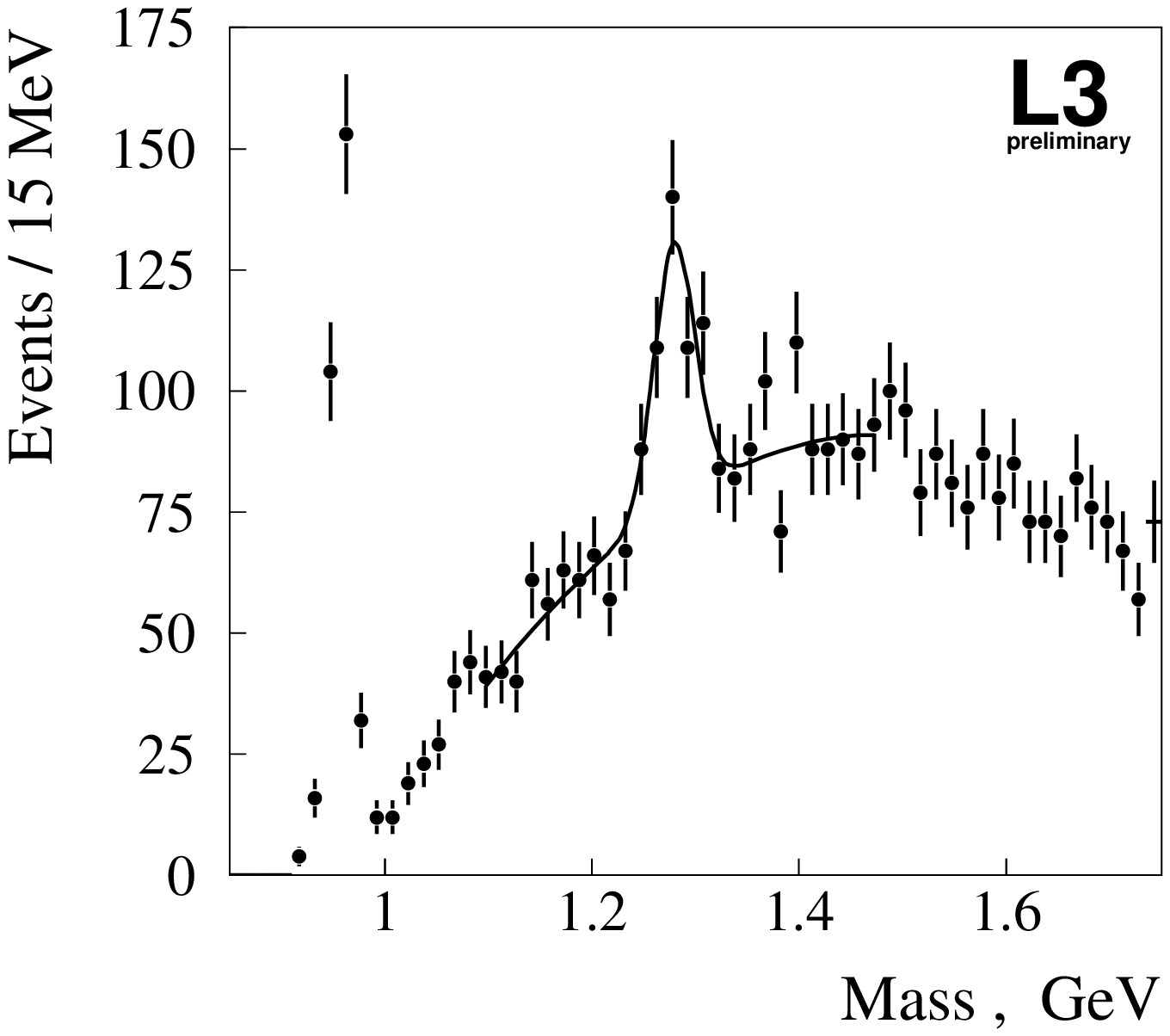}
\caption{The $\eta\pi^+\pi^-$ mass spectrum for $Q^2<$0.02 GeV$^2$(left) and
$Q^2>$0.02 GeV$^2$(right).}
\label{fig:etapipi}
\end{figure}

 Interesting new preliminary results on the
K$_s^0$K$^\pm\pi^\mp$ and the $\eta\pi^+\pi^-$ final states are obtained by L3~\cite{Igor}.
The K$_s^0$K$^\pm\pi^\mp$ mass spectrum is studied as a function of $Q^2$ (fig.~\ref{fig:k0kp}).
A prominent signal is present at 1470 MeV at low and at high $Q^2$. At very high $Q^2$
another signal appears around 1300 MeV due to the formation of the f$_1$(1285).
The study of the cross section as a function of $Q^2$ in the 1470 MeV region
reveals that both the 0$^{-+}$ and 1$^{++}$ waves are needed to fit the data. 
The 0$^{-+}$ wave is due to the
formation of the $\eta(1440)$ and largely dominates at low $Q^2$ while at high $Q^2$ the formation
of the f$_1$(1420) is found to be dominant. The value
$\Gamma_{\gamma\gamma}(\eta(1440))\times \mbox{BR}(\eta(1440)\rightarrow \mbox{K}\bar{\mbox{K}}\pi)$ 
= 234 $\pm$ 55  $\pm$  17 eV
is obtained by using data at low $Q^2$. 
This first observation of the $\eta(1440)$ in untagged two-photon
collisions disfavours its interpretation as the 0$^{-+}$ glueball in agreement with the lattice QCD
calculations. The $\eta(1440)$ can therefore be interpreted as a radial excitation~\cite{Anisovitch-1}.
The $\eta\pi^+\pi^-$ final state shows no evidence for the
formation of the $\eta(1440)$ at low and at
high $Q^2$ (fig.~\ref{fig:etapipi}). A prominent signal due to the formation
of the $\eta'$(958) is present in the two spectra
while the f$_1$(1285) is visible only at high  $Q^2$.
The upper limits 
$\Gamma_{\gamma\gamma}(\eta(1440))\times\mbox{BR}(\eta(1440)\rightarrow
\eta\pi\pi)<$ 88 eV and
$\Gamma_{\gamma\gamma}(\eta(1295))\times\mbox{BR}(\eta(1295)\rightarrow
\eta\pi\pi)<$ 61 eV at 90\% C.L. are obtained.

\section{Glueball searches in the K$_s^0$K$_s^0$ and $\pi^+\pi^-$ final states }

\begin{figure}[t]
\includegraphics[width=0.45\textwidth]{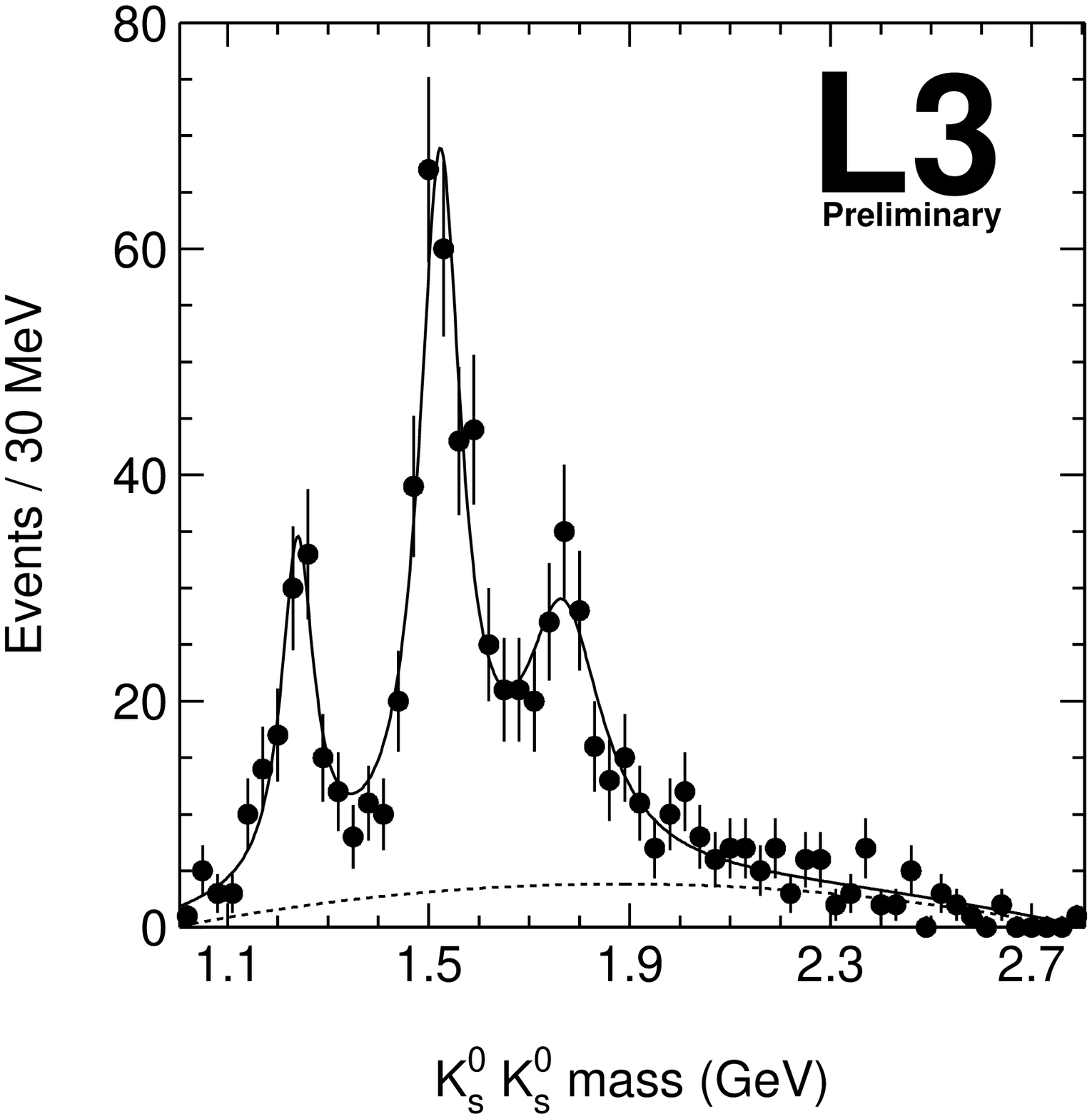}\hspace*{.3cm}
\includegraphics[width=0.45\textwidth]{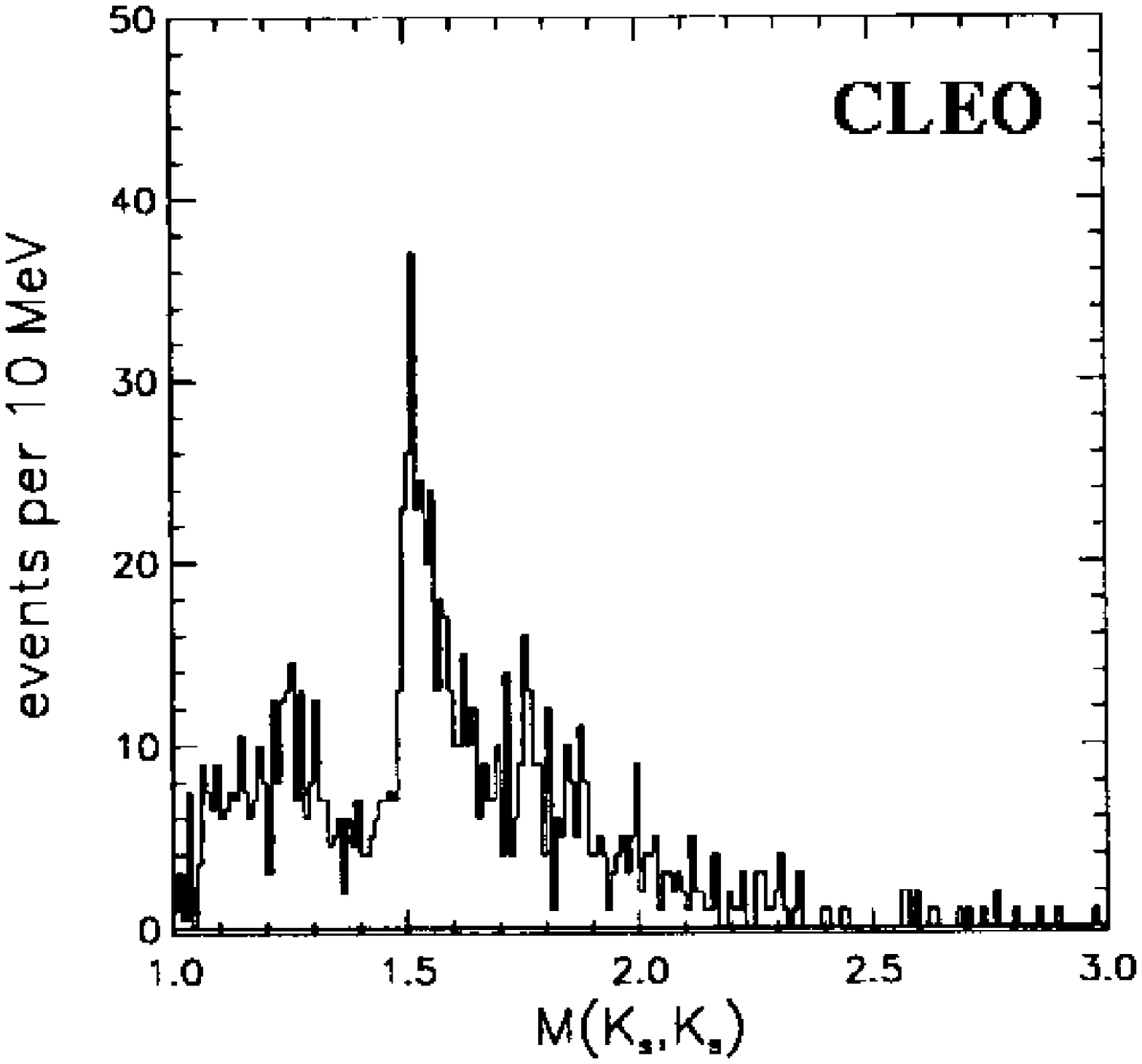}
\caption{The K$_s^0$K$_s^0$ mass spectra measured by L3 (left) and CLEO (right).}
\label{fig:k0k0}
\end{figure}

A study of the reaction $\gamma\gamma\rightarrow$ K$^0_s$K$^0_s$ is performed 
by L3~\cite{Saverio-Frascati}~\cite{Saverio}.
The mass spectrum is shown in fig.~\ref{fig:k0k0}(left).
The 1100-1400 MeV mass region
shows destructive f$_2$(1270) -- a$_2$(1320) interference 
in agreement with theoretical predictions~\cite{Lipkin}.
The spectrum is dominated by the formation of the f$_2\,\!\!\!'$(1525) tensor meson
in helicity 2 state as clearly shown by the angular distribution
in the $\kos\kos$ center of mass.
The preliminary value 
$\Gamma_{\gamma\gamma}(f_2'(1525))\times \mbox{BR}(f_2'(1525)\rightarrow
\mbox{K}\bar{\mbox{K}})$= 0.076 $\pm$ 0.006 $\pm$ 0.011 keV is obtained.  
A clear signal is present in the 1750 MeV mass region  due to the
formation of the f$_J$(1710). The presence of a 0$^{++}$ $s\bar{s}$ meson 
would support the glueball interpretation
of the f$_0$(1500)~\cite{Amsler}. The study of the angular distribution in the 1750 MeV mass region
favours the presence of a 2$^{++}$, helicity 2 wave. This is consistent with the interpretation of the
f$_J$(1710) as a radial recurrence of the f$_2\,\!\!\!'$(1525)~\cite{Munz}.
The presence of a 0$^{++}$ wave cannot however be excluded.
The BES Collaboration~\cite{BESKPKM} reported the presence of both 2$^{++}$ and 0$^{++}$
waves in the 1750 MeV region in K$^+$K$^-$ in the
reaction $\epem\ra\mbox{J}/\psi\ra\mbox{K}^+\mbox{K}^-\gamma$. 
No signal for the formation of
the $\xi$(2230)~\cite{xi} tensor glueball candidate is observed. 
The upper limit $\Gamma_{\gamma\gamma}(\xi(2230))\times $BR$(\xi(2230)\ra\kos\kos)<1.4$ eV 
at 95\% C.L. is obtained. The stickiness is found to be $S_{\xi(2230)} >$ 73 at 95\% C.L.

 The  $\xi(2230)$ is searched by CLEO in the K$_s^0$K$_s^0$~\cite{CLEO-k0k0} and 
$\pi^+\pi^-$~\cite{CLEO-pipi} final states. The K$_s^0$K$_s^0$ 
mass spectrum (fig.~\ref{fig:k0k0}(right)) shows
similar features respect to the L3 data. The upper limits
$\Gamma_{\gamma\gamma}(\xi(2230))\times\mbox{BR}(\xi(2230)\ra\kos\kos)<$ 1.3 eV
and
$\Gamma_{\gamma\gamma}(\xi(2230))\times\mbox{BR}(\xi(2230)\rightarrow \pi^+\pi^-)<$ 2.5 eV 
at 95\% C.L. are obtained. Combining these two results the stickiness is found to be
$S_{\xi(2230)} >$ 102 at 95\% C.L.
The very large lower limits for $S_{\xi(2230)}$
obtained by CLEO and L3 give a strong support to the interpretation
of the $\xi$(2230) as the tensor glueball. A confirmation of its existence 
in gluon rich environments becomes now very important.

 The $\pi^+\pi^-$ final state is studied by ALEPH~\cite{ALEPH-pipi}. The mass spectrum (fig.~\ref{fig:pipialeph})
shows a signal due to the formation of the f$_2$(1270) tensor meson. No other signals are present.
Assuming the f$_0$(1500) and the f$_J$(1710) to be scalars, the upper limits
$\Gamma_{\gamma\gamma}(f_0(1500))\times\mbox{BR}(f_0(1500)\rightarrow
\pi^+\pi^-)<$ 310 eV 
and
$\Gamma_{\gamma\gamma}(f_J(1710))\times\mbox{BR}(f_J(1710)\rightarrow
\pi^+\pi^-)<$ 550 eV at 95\% C.L.
are obtained. Interference effects with the $\pi^+\pi^-$ continuum are not taken into account.
According to A.V.Anisovitch et al.~\cite{Anisovitch}, interference 
with the $\pi^+\pi^-$ continuum should make the f$_0$(1500) appear as a dip.

\section{Charmonium formation }

\begin{figure}[t]
\begin{minipage}[t]{60mm}
\mbox{\epsfig{file=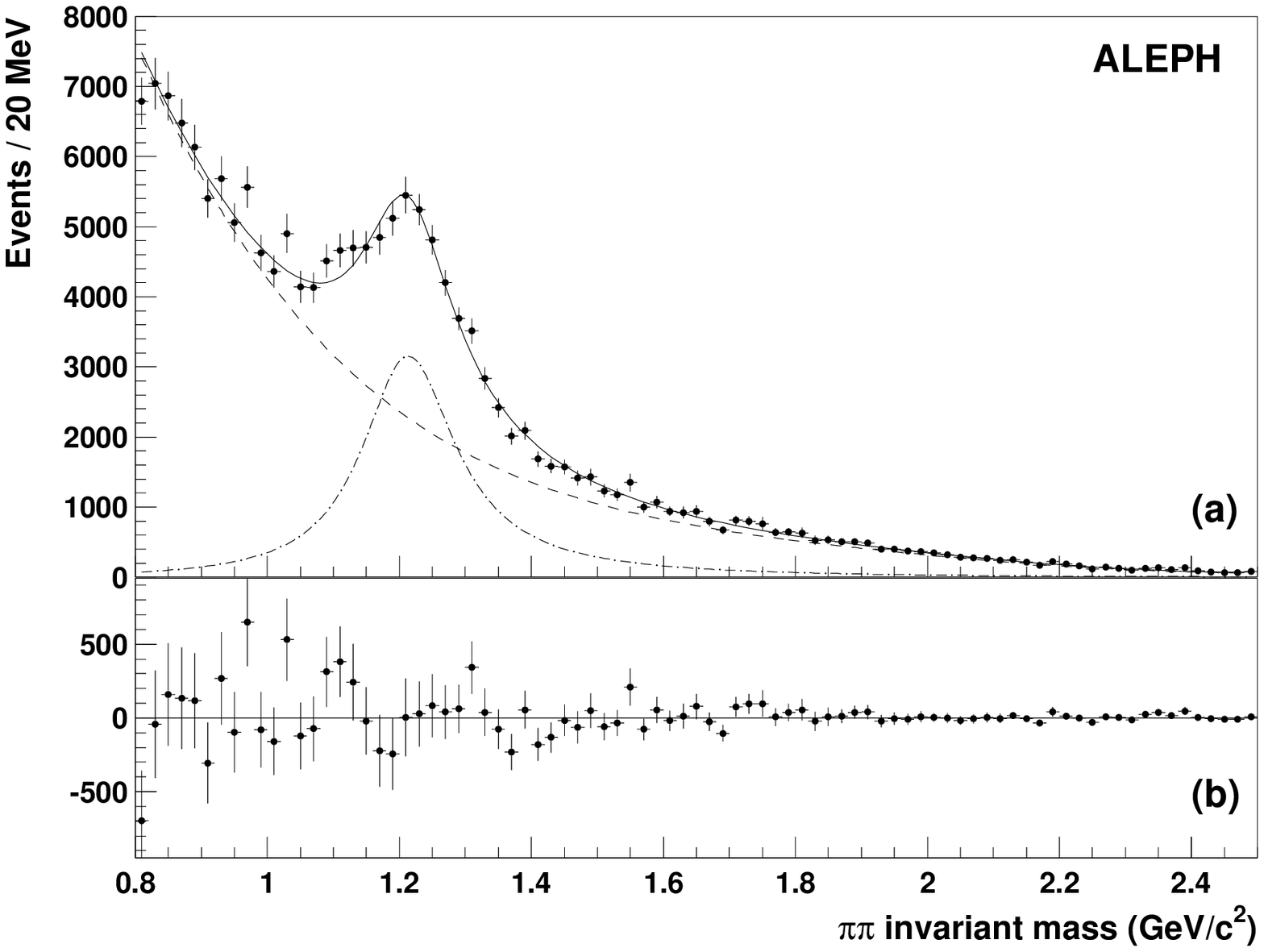,width=6.cm}}
\vspace*{-.4cm}
\caption{The $\pi^+\pi^-$ mass spectrum: the only signal is due to
the formation of the f$_2$(1270).
}
\label{fig:pipialeph}
\end{minipage}
\hspace{\fill}
\begin{minipage}[t]{60mm}
\mbox{\epsfig{file=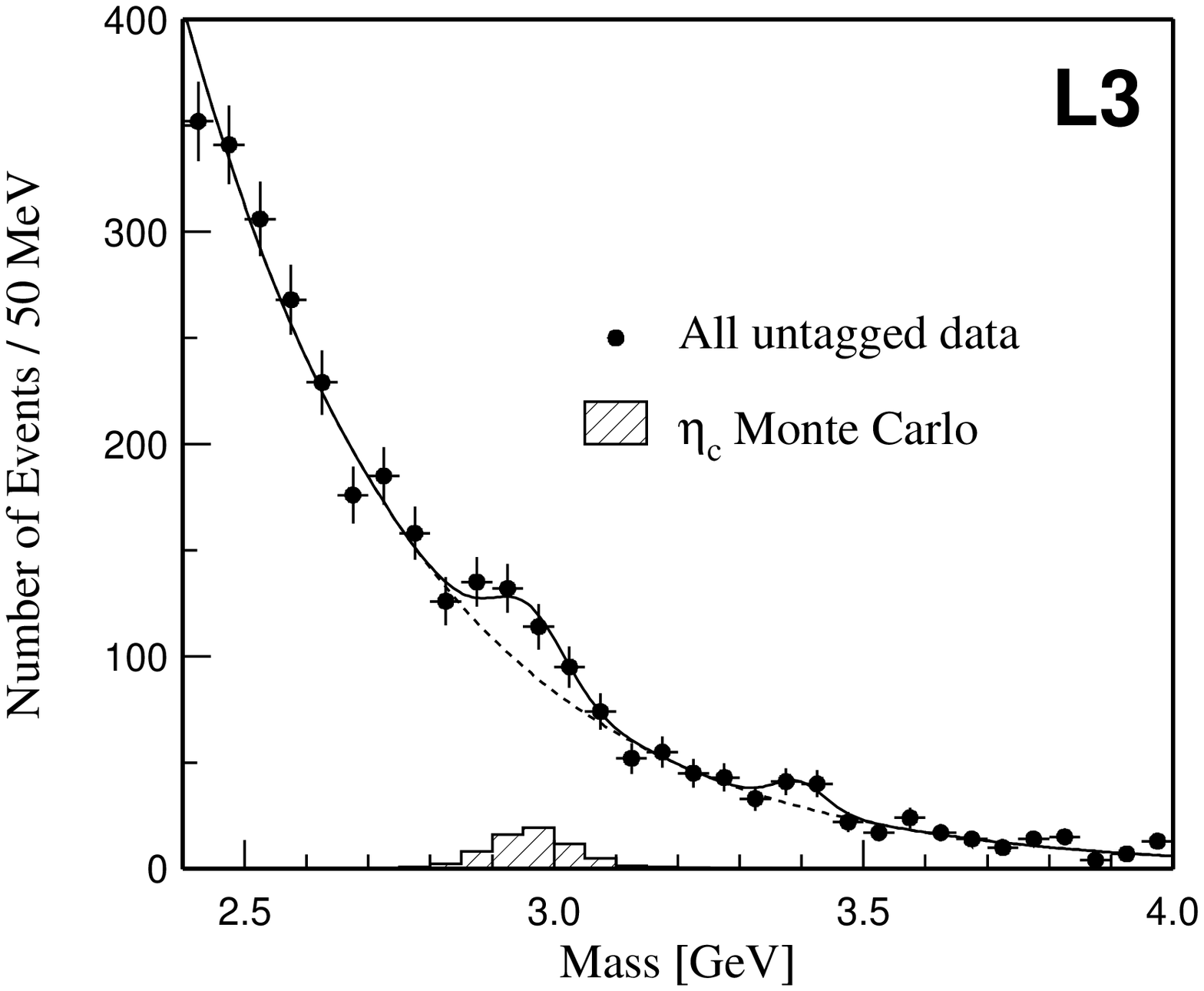,width=6.cm}}
\vspace*{-.4cm}
\caption{The mass spectrum of the sum of nine different final states.
The signals of the $\eta_c(2980)$ and the $\chi_{c0}(3415)$ are visible.
}
\label{fig:etac}
\end{minipage}
\vspace*{-.4cm}
\end{figure}

The formation of the $\eta_c(2980)$ is studied by L3~\cite{L3-etac}.
Since the $\eta_c$ decays in many different final states with small branching fractions,
the simultaneous study of several decay channels is mandatory. The mass spectrum shown
in fig.~\ref {fig:etac} is obtained by summing nine different final states. The value 
$\Gamma_{\gamma\gamma}(\eta_c)$ = 6.9 $\pm$ 1.7 (stat.)  $\pm$  0.8
(sys.)$\pm$ 2.0 (BR) keV 
is measured. Despite the limited statistics, the study of the formation 
of the $\eta_c(2980)$ as a function of $Q^2$ allows
to exclude a VDM $\rho$ pole transition form factor. Data are consistent with a J/$\psi$ VDM pole
form factor, as expected. 

\begin{figure}[t]
\begin{minipage}[t]{60mm}
\mbox{\epsfig{file=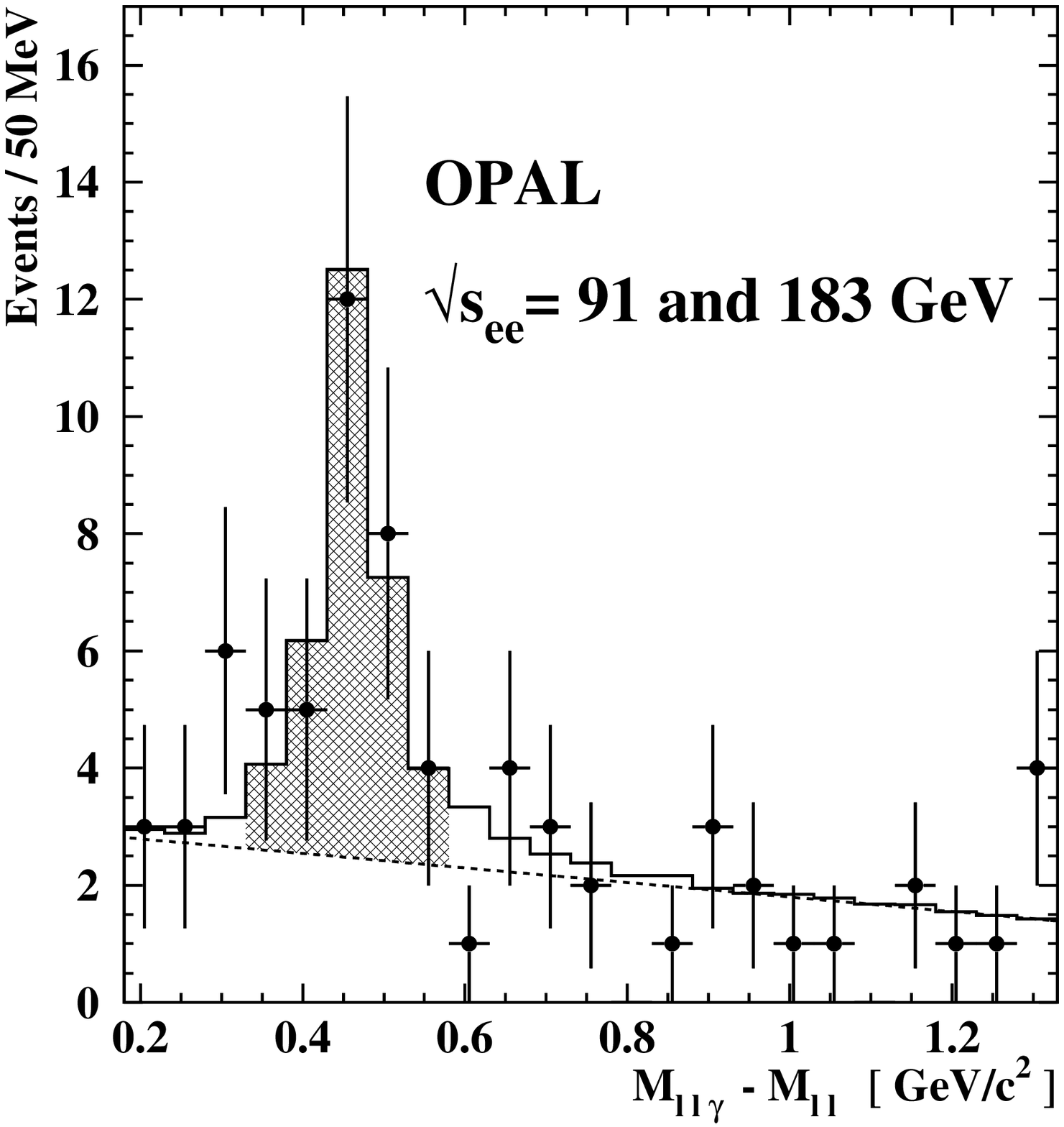,width=6.cm}}
\vspace*{-.4cm}
\caption{The signal of the formation of the $\chi_{c2}(3555)$ from the mass
difference m$(l^+l^-\gamma)$-m$(l^+l^-)$. 
}
\label{fig:chic_opal}
\end{minipage}
\hspace{\fill}
\begin{minipage}[t]{60mm}
\mbox{\epsfig{file=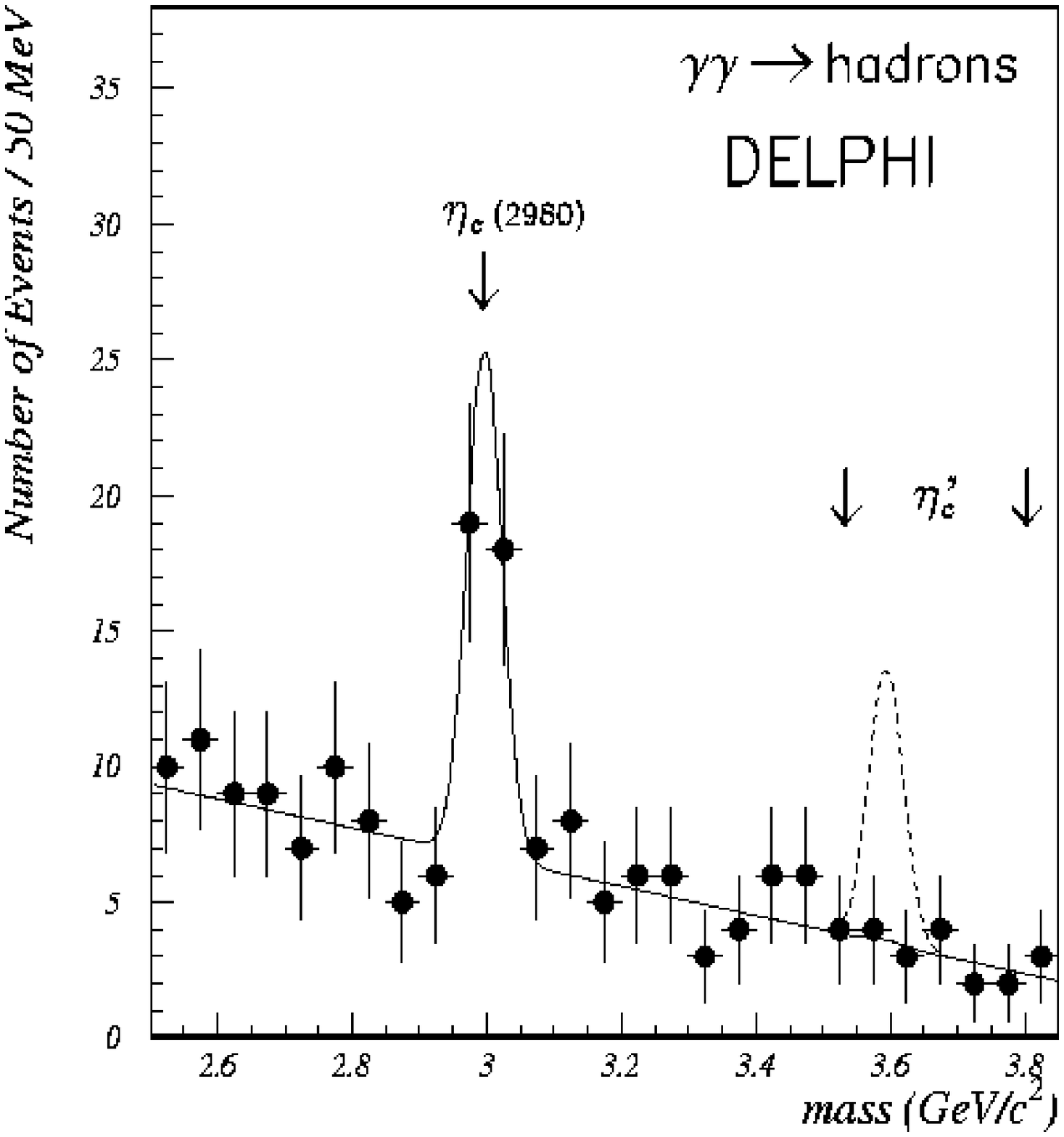,width=6.cm}}
\vspace*{-.4cm}
\caption{The mass spectrum of the sum of five different decay channels.
}
\label{fig:etacp_delphi}
\end{minipage}
\vspace*{-.4cm}
\end{figure}

From the reaction  
$\gamma\gamma\rightarrow\chi_{c2}(3555)\rightarrow J/\psi \gamma\rightarrow l^+l^-\gamma$ 
with $l=e,\mu$, the two-photon width of the $\chi_{c2}$ is measured by OPAL\cite{OPAL-chic}.
The signal is seen in the distribution of the mass
difference m$(l^+l^-\gamma)$ -- m$(l^+l^-)$ when m$(l^+l^-)$ is compatible with the mass
of the J$/\psi$ (fig.~\ref{fig:chic_opal}). The value
$\Gamma_{\gamma\gamma}(\chi_{c2})$ = 1.76 $\pm$ 0.47 (stat.)  $\pm$ 
0.37 (sys.)$\pm$ 0.15 (BR) keV is obtained.
The value 
$\Gamma_{\gamma\gamma}(\chi_{c2})$ = 1.02 $\pm$ 0.40 (stat.)  $\pm$ 
0.15 (sys.)$\pm$ 0.09 (BR) keV
is measured by L3~\cite{L3-chic} using the same method.

 The measurements of the two-photon width of the $\eta_c$ performed in two-photon collisions
are in good agreement with the ones obtained in $p\bar{p}$ annihilations~\cite{PDG}.
For the $\chi_{c2}$ the agreement is not good and the two-photon measurements 
are significantly higher than the value 
$\Gamma_{\gamma\gamma}(\chi_{c2})$ = 0.31 $\pm$ 0.05 $\pm$ 0.04 keV
measured by E835 at Fermilab~\cite{Stancari} in $p\bar{p}$ annihilations.
This value is in agreement with a previous measurement by E760~\cite{E760-chic}.
The reason for this is not known but it is interesting to
remark that all the two-photon
measurements are performed by using the same final state and the same
experimental method.
 
 No signal for the formation of the $\eta_c'$ is observed at LEP. 
Five different decay channels are examined by DELPHI~\cite{DELPHI-etacp}
as shown in fig.~\ref{fig:etacp_delphi}. The formation of
the $\eta_c(2980)$ is clearly observed while no signal is present in the 
$\eta_c'$ mass region. The upper limit  
$\frac{\Gamma_{\gamma\gamma}(\eta_c')}{\Gamma_{\gamma\gamma}(\eta_c)}<0.34$ 
at 90\% C.L.
is obtained. 
The upper limit
$\Gamma_{\gamma\gamma}(\eta_c')<2.0$ keV at 95\% C.L.
is obtained by L3~\cite{L3-etac} using nine different decay modes.

\section{Conclusions }

\begin{table}[t]
\begin{tabular}{|l|c|c|c|c|c|}
\hline
Resonance  & Experiment & Final state  & J$^{PC}$& $\Gamma_{\gamma\gamma}$ & Ref. \\
\hline 
 $\eta'$(958)         & L3             & $\pi^+ \pi^- \gamma$ & $ 0^{-+}   $ & 4.17$\pm$0.10$\pm$0.27 keV & ~\cite{L3-etap}\\
 $a_2$(1320)          & L3             & $\pi^+ \pi^- \pi^0 $ & $ 2^{++}   $ & 0.98$\pm$0.05$\pm$0.09 keV & ~\cite{L3-a2} \\
 $f_2^{'}$(1525)      & L3             & K$^0_s$K$^0_s$       & $ 2^{++}   $ & 0.085$\pm$0.007$\pm$0.012 keV & ~\cite{Saverio}\\ \hline
 $\eta _{c}$(2980)    & L3	       & 9 chan.  	      & $ 0^{-+}   $ & 6.9$\pm$1.7.$\pm$0.8  keV & ~\cite{L3-etac}\\
 $\eta _{c}'$         & L3	       & 9 chan.  	      & $ 0^{-+}   $ & $<2.0$ keV  & ~\cite{L3-etac}\\
 $\chi_{c2}$(3555)    & L3	       & $ l^+ l^- \gamma $   & $ 2^{++}   $ & 1.02$\pm$0.40$\pm$0.15 keV & ~\cite{L3-chic}\\ 
 $\chi_{c2}$(3555)    & OPAL	       & $ l^+ l^- \gamma $   & $ 2^{++}   $ & 1.76$\pm$0.47$\pm$0.37 keV & ~\cite{OPAL-chic}\\ \hline
 $\eta$(1440)         & L3	       & K$^0_s$K$^\pm\pi^\mp$& $ 0^{-+}   $ & 234$^{\dag}\pm$55$\pm$17 eV& ~\cite{Igor}\\
 f$_J$(1710)	      & L3	       & K$^0_s$K$^0_s$       & $ (?)^{++} $ &                   & ~\cite{Saverio}\\
 $a_2'$(1752)         & L3	       & $\pi^+ \pi^- \pi^0 $ & $ 2^{++}   $ & 0.29$^{\dag}\pm$0.04$\pm$0.02 keV & ~\cite{L3-a2}\\ \hline
 f$_0$(1500)	      & ALEPH	       & $\pi^+ \pi^-$        & $ 0^{++}   $ & $<310^{\dag}$ eV & ~\cite{ALEPH-pipi}\\
 f$_0$(1710)	      & ALEPH	       & $\pi^+ \pi^-$        & $ 0^{++}   $ & $<550^{\dag}$ eV & ~\cite{ALEPH-pipi}\\
 $\xi$(2230)	      & CLEO	       & $\pi^+ \pi^-$        & $ 2^{++}   $ & $<2.5^{\dag}$ eV& ~\cite{CLEO-pipi}\\
 $\xi$(2230)	      & CLEO	       & K$^0_s$K$^0_s$       & $ 2^{++}   $ & $<1.3^{\dag}$ eV& ~\cite{CLEO-k0k0}\\
 $\xi$(2230)	      & L3	       & K$^0_s$K$^0_s$       & $ 2^{++}   $ & $<1.4^{\dag}$ eV& ~\cite{Saverio}\\ \hline
\end{tabular}
\caption{The most recent results on the two-photon width of mesons, charmonia, radial excitations and glueball
candidates. 
($\dag$ the value is given times the decay branching ratio)}
\label{summary}
\end{table}

 A remarkable progress on the study of resonance formation in
two-photon collisions has been achieved in the last few years.
Data from  the LEP collider at CERN and CESR at Cornell allowed to
improve significantly the precision on the two-photon widths of
several resonances, to study the transition form factors, to identify
some radial excitations and to search for glueball candidates.
All these results are summarised in Table~\ref{summary}. They 
represent an important contribution to meson
spectroscopy and glueball searches.

\section{Acknowledgements}

I would like to acknowledge the two-photon physics groups
of the ALEPH, CLEO, DELPHI, L3 and OPAL collaborations.
I would like to thank M.N. Focacci-Kienzle,
J.H. Field, M. Wadhwa, I. Vodopianov, A. Buijs, H. P. Paar, C. Amsler,
L. Montanet,
U. Gastaldi and P. Minkowski for the very constructive discussions and suggestions.
I would like to express my gratitude to B. Monteleoni, recently deceased.

%

\end{document}